\documentstyle[twoside]{article}

\catcode`\@=11
\long\def\@makefntext#1{
\protect\noindent \hbox to 3.2pt {\hskip-.9pt
$^{{\eightrm\@thefnmark}}$\hfil}#1\hfill}		

\def\thefootnote{\fnsymbol{footnote}}
\def\@makefnmark{\hbox to 0pt{$^{\@thefnmark}$\hss}}	

\def\ps@myheadings{\let\@mkboth\@gobbletwo
\def\@oddhead{\hbox{}
\rightmark\hfil\eightrm\thepage}
\def\@oddfoot{}\def\@evenhead{\eightrm\thepage\hfil
\leftmark\hbox{}}\def\@evenfoot{}
\def\sectionmark##1{}\def\subsectionmark##1{}}

\oddsidemargin=\evensidemargin
\renewcommand{\thefootnote}{\fnsymbol{footnote}}

\newcounter{sectionc}
\newcounter{subsectionc}
\newcounter{subsubsectionc}
\renewcommand{\section}[1] {\vspace{12pt}\addtocounter{sectionc}{1}
\setcounter{subsectionc}{0}\setcounter{subsubsectionc}{0}\noindent
	{\tenbf\thesectionc. #1}\par\vspace{5pt}}
\renewcommand{\subsection}[1] {\vspace{12pt}
\addtocounter{subsectionc}{1}\setcounter{subsubsectionc}{0}\noindent
	{\bf\thesectionc.\thesubsectionc.
        {\kern1pt \bfit #1}}\par\vspace{5pt}}
\renewcommand{\subsubsection}[1] {\vspace{12pt}
\addtocounter{subsubsectionc}{1}\noindent
        {\tenrm\thesectionc.\thesubsectionc.\thesubsubsectionc.
	{\kern1pt \tenit #1}}\par\vspace{5pt}}

\newcounter{appendixc}
\newcounter{subappendixc}[appendixc]
\newcounter{subsubappendixc}[subappendixc]
\renewcommand{\thesubappendixc}{\Alph{appendixc}.
        \arabic{subappendixc}}
\renewcommand{\thesubsubappendixc}{\Alph{appendixc}.
        \arabic{subappendixc}.\arabic{subsubappendixc}}

\renewcommand{\appendix}[1] {\vspace{12pt}
        \refstepcounter{appendixc}
        \setcounter{figure}{0}
        \setcounter{table}{0}
        \setcounter{lemma}{0}
        \setcounter{theorem}{0}
        \setcounter{corollary}{0}
        \setcounter{definition}{0}
        \setcounter{equation}{0}
        \renewcommand{\thefigure}{\Alph{appendixc}.\arabic{figure}}
        \renewcommand{\thetable}{\Alph{appendixc}.\arabic{table}}
        \renewcommand{\theappendixc}{\Alph{appendixc}}
        \renewcommand{\thelemma}{\Alph{appendixc}.\arabic{lemma}}
        \renewcommand{\thetheorem}{\Alph{appendixc}.\arabic{theorem}}
        \renewcommand{\thedefinition}{\Alph{appendixc}.
         \arabic{definition}}
        \renewcommand{\thecorollary}{\Alph{appendixc}.
         \arabic{corollary}}
        \renewcommand{\theequation}{\Alph{appendixc}.
         \arabic{equation}}
        \noindent{\tenbf Appendix \theappendixc #1}\par\vspace{5pt}}
\newcommand{\subappendix}[1] {\vspace{12pt}
        \refstepcounter{subappendixc}
        \noindent{\bf Appendix \thesubappendixc. {\kern1pt \bfit #1}}
	\par\vspace{5pt}}
\newcommand{\subsubappendix}[1] {\vspace{12pt}
        \refstepcounter{subsubappendixc}
        \noindent{\rm Appendix \thesubsubappendixc.
        {\kern1pt \tenit #1}}\par\vspace{5pt}}

\newcommand{\textlineskip}{\baselineskip=13pt}
\newcommand{\smalllineskip}{\baselineskip=10pt}

\def\eightcirc{
\begin{picture}(0,0)
\put(4.4,1.8){\circle{6.5}}
\end{picture}}
\def\eightcopyright{\eightcirc\kern2.7pt\hbox{\eightrm c}}

\def\abstracts#1#2#3{{
	\centering{\begin{minipage}{4.5in}\baselineskip=10pt
        \footnotesize
	\parindent=0pt #1\par
	\parindent=15pt #2\par
	\parindent=15pt #3
	\end{minipage}}\par}}

\renewenvironment{thebibliography}[1]
	{\frenchspacing
	 \ninerm\baselineskip=11pt
	 \begin{list}{\arabic{enumi}.}
	{\usecounter{enumi}\setlength{\parsep}{0pt}
	 \setlength{\leftmargin 12.7pt}{\rightmargin 0pt}
	 \setlength{\itemsep}{0pt} \settowidth
	{\labelwidth}{#1.}\sloppy}}{\end{list}}

\newcounter{itemlistc}
\newcounter{romanlistc}
\newcounter{alphlistc}
\newcounter{arabiclistc}

\newcommand{\fcaption}[1]{
        \refstepcounter{figure}
        \setbox\@tempboxa = \hbox{\footnotesize Fig.~\thefigure. #1}
        \ifdim \wd\@tempboxa > 5in
           {\begin{center}
        \parbox{5in}{\footnotesize\smalllineskip Fig.~\thefigure. #1}
            \end{center}}
        \else
             {\begin{center}
             {\footnotesize Fig.~\thefigure. #1}
              \end{center}}
        \fi}

\newcommand{\tcaption}[1]{
        \refstepcounter{table}
        \setbox\@tempboxa = \hbox{\footnotesize Table~\thetable. #1}
        \ifdim \wd\@tempboxa > 5in
           {\begin{center}
        \parbox{5in}{\footnotesize\smalllineskip Table~\thetable. #1}
            \end{center}}
        \else
             {\begin{center}
             {\footnotesize Table~\thetable. #1}
              \end{center}}
        \fi}

\def\@citex[#1]#2{\if@filesw\immediate\write\@auxout
	{\string\citation{#2}}\fi
\def\@citea{}\@cite{\@for\@citeb:=#2\do
	{\@citea\def\@citea{,}\@ifundefined
	{b@\@citeb}{{\bf ?}\@warning
	{Citation `\@citeb' on page \thepage \space undefined}}
	{\csname b@\@citeb\endcsname}}}{#1}}

\newif\if@cghi
\def\cite{\@cghitrue\@ifnextchar [{\@tempswatrue
	\@citex}{\@tempswafalse\@citex[]}}
\def\citelow{\@cghifalse\@ifnextchar [{\@tempswatrue
	\@citex}{\@tempswafalse\@citex[]}}
\def\@cite#1#2{{$\null^{#1}$\if@tempswa\typeout
	{IJCGA warning: optional citation argument
	ignored: `#2'} \fi}}

\def\pmb#1{\setbox0=\hbox{#1}
	\kern-.025em\copy0\kern-\wd0
	\kern.05em\copy0\kern-\wd0
	\kern-.025em\raise.0433em\box0}

\def\fnt#1#2{\footnotetext{\kern-.3em
	{$^{\mbox{\scriptsize #1}}$}{#2}}}

\def\fpage#1{\begingroup
\voffset=.3in
\thispagestyle{empty}\begin{table}[b]\centerline{\footnotesize #1}
	\end{table}\endgroup}

\font\tenrm=cmr10
\font\tenit=cmti10
\font\tenbf=cmbx10
\font\bfit=cmbxti10 at 10pt
\font\ninerm=cmr9

\font\eightrm=cmr8

\def\qed{\hbox{${\vcenter{\vbox{		
   \hrule height 0.4pt\hbox{\vrule width 0.4pt height 6pt
   \kern5pt\vrule width 0.4pt}\hrule height 0.4pt}}}$}}

\renewcommand{\thefootnote}{\fnsymbol{footnote}}

\input epsf
\global\arraycolsep=2pt
\def\spose#1{\hbox to 0pt{#1\hss}}
\def\lsim{\mathrel{\spose{\lower 3pt\hbox{$\mathchar"218$}}
 \raise 2.0pt\hbox{$\mathchar"13C$}}}
\def\gsim{\mathrel{\spose{\lower 3pt\hbox{$\mathchar"218$}}
 \raise 2.0pt\hbox{$\mathchar"13E$}}}

\renewcommand{\theequation}{\thesection.\arabic{equation}}
\def\laq{\raise 0.4ex\hbox{$<$}\kern -0.8em\lower 0.62
ex\hbox{$\sim$}}
\def\gaq{\raise 0.4ex\hbox{$>$}\kern -0.7em\lower 0.62
ex\hbox{$\sim$}}

\def\beq{\begin{equation}}
\def\eeq{\end{equation}}
\def\bea{\begin{eqnarray}}
\def\eea{\end{eqnarray}}

\def \pa {\partial}
\def \ra {\rightarrow}

\def \fb {\overline \phi}
\def \fbp {\dot{\fb}}

\def \pr {\prime}

\def \H {{a^\prime \over a}}

\def \la {\lambda}

\def \La {\Lambda}
\def \Da {\Delta}
\def \b {\beta}
\def \a {\alpha}
\def \ap {\alpha^{\prime}}

\def \Ga {\Gamma}
\def \ga {\gamma}

\def \da {\delta}
\def \ep {\epsilon}
\def \r {\rho}
\def \om {\omega}
\def \Om {\Omega}
\def \noi {\noindent}

\begin{document}

\begin{titlepage}

\begin{flushright}
CERN-TH/96-330\\
gr-qc/9611059
\end{flushright}

\vspace{2 cm}

\begin{center}
\Large\bf Relic Dilatons in String Cosmology
\end{center}

\vspace{1.5cm}

\begin{center}
M. Gasperini\\
{\sl Theory Division, CERN, CH-1211 Geneva 23, Switzerland}\\
and\\
{\sl Dipartimento di Fisica Teorica, Universit\`a di Torino,}\\
{\sl Via P. Giuria 1, 10125 Turin, Italy}
\end{center}

\vspace{1.5cm}

\begin{abstract}
\noi
The allowed mass windows for a cosmic background of relic dilatons
are estimated in the context of the pre-big bang scenario. The
dilatons are produced from the quantum fluctuations of the vacuum, 
and the extension of the windows is controlled by the string mass
scale. The possible relaxation of phenomenological bounds due to an
intermediate stage of reheating is discussed. Even without such a
relaxation, the allowed range of masses includes a light sector in  which
the dilatons are not yet decayed, and could provide
the dominant contribution to the present large scale density.

\end{abstract}

\vspace{1.5cm}
\begin{center}
To appear in \\
{\sl Proc. of the 12th Italian Conf. on General Relativity and
Gravitational Physics} \\
Rome, September 1996\\ 
ed. by M. Bassan et al., to be published by 
World Scientific (Singapore, 1996)
\end{center}
 \vspace{1.5cm}
\vfill
\begin{flushleft}
CERN-TH/96-330\\
November 1996 
\end{flushleft}

\end{titlepage}

\thispagestyle{empty}
\vbox{}
\newpage

\normalsize\textlineskip
\thispagestyle{empty}
\setcounter{page}{1}


\vspace*{0.18truein}

\fpage{1}

\centerline{\bf RELIC DILATONS IN STRING COSMOLOGY}
\vspace*{0.27truein}

\centerline{\footnotesize MAURIZIO GASPERINI}
\vspace*{0.015truein}
\centerline{\footnotesize\it Theory Division, CERN,
 CH-1211 Geneva 23, Switzerland}
\baselineskip=10pt
\centerline{\footnotesize and {\it Dipartimento di Fisica Teorica, 
Universit\`a di Torino, Turin, Italy}}
\vspace*{0.225truein}

\vspace*{0.11truein}
\abstracts
{The allowed mass windows for a cosmic background of relic dilatons
are estimated in the context of the pre-big bang scenario. The
dilatons are produced from the quantum fluctuations of the vacuum, 
and the extension of the windows is controlled by the string mass
scale. The possible relaxation of phenomenological bounds due to an
intermediate stage of reheating is discussed. Even without such a
relaxation, the allowed range of masses includes a light sector in  which
the dilatons are not yet decayed, and could provide
the dominant contribution to the present large scale density.}{}{}
\vspace*{1pt}\textlineskip

\textheight=7.8truein
\setcounter{footnote}{0}
\renewcommand{\thefootnote}{\alph{footnote}}

\vspace*{0.3truein}

\renewcommand{\theequation}{1.\arabic{equation}}
\setcounter{equation}{0}
\section{Introduction}
\label{sec:1}
\noindent
String theory has recently motivated the study of a cosmological
scenario in which the Universe starts evolving from the string
perturbative vacuum, namely from a cold and empty state with flat
metric and vanishing gauge coupling. Because of the instability of this
vacuum the Universe is necessarily driven, in a finite amount of 
time, to a state with high curvature and strong coupling, where the
back-reaction of the quantum fluctuations becomes important, and the
Universe eventually becomes hot and radiation-dominated as in the
standard scenario. The big bang, in such a context, is no longer the
starting point of the cosmological evolution, but only the intermediate
stage corresponding to the transition from the high-curvature string
phase to the standard radiation era. It thus seems appropriate to call
``pre-big bang"\cite{1,1a,1b} the whole cosmological epoch
describing the evolution from the vacuum to the beginning of
radiation-dominance, characterized by shrinking event horizons,
growing curvature and string coupling, and naturally motivated by the
duality symmetries of the string effective action.

By assuming, for simplicity, a non-trivial background configuration for
the metric and the dilaton field only, the effective action\cite{2}
governing the dynamics of the pre-big bang cosmological scenario can
be written, in the string frame:
\bea
S=&-&\int d^{d+1}x\sqrt{|g|}e^{-\phi}\left[R+(\nabla
\phi)^2- {\ap\over 4}\left( R^2_{\mu\nu\a\b} -4R^2_{\mu\nu}+R^2
-(\nabla\phi)^4+ ...\right) +\right. \nonumber \\
 &+& \left. V(\phi)\right] +
{\rm loops} (g_s) .
\label{11}
\eea
Here $\phi$ is the dilaton, $V$ is a possible non-perturbative dilaton
potential, $g_s(t)=e^{\phi/2}$ is the field-dependent (and thus
time-dependent) string coupling, and the dots stand for
higher-derivative terms, whose contribution to the effective action is
controlled by the fundamental string mass parameter:
\beq
\ap \equiv \la_s^2 \equiv M_s^{-2} .
\label{12}
\eeq
For an isotropic and spatially flat $d$-dimensional background,
\beq
g_{\mu\nu}={\rm diag}\left(1, -a^2(t)\da_{ij}\right) , ~~~ \phi=
\phi(t), ~~~\fbp=\dot\phi-dH, ~~~ H=\dot a /a,
\label{13}
\eeq
the cosmological evolution determined by the action (\ref{11}) can be
schematically illustrated as in Fig. 1, in the two-dimensional space
spanned by the convenient dynamical variables $\{\fbp, \sqrt d H\}$.

\begin{figure}[htb]
   \epsfxsize=11cm
   \centerline{\epsfbox{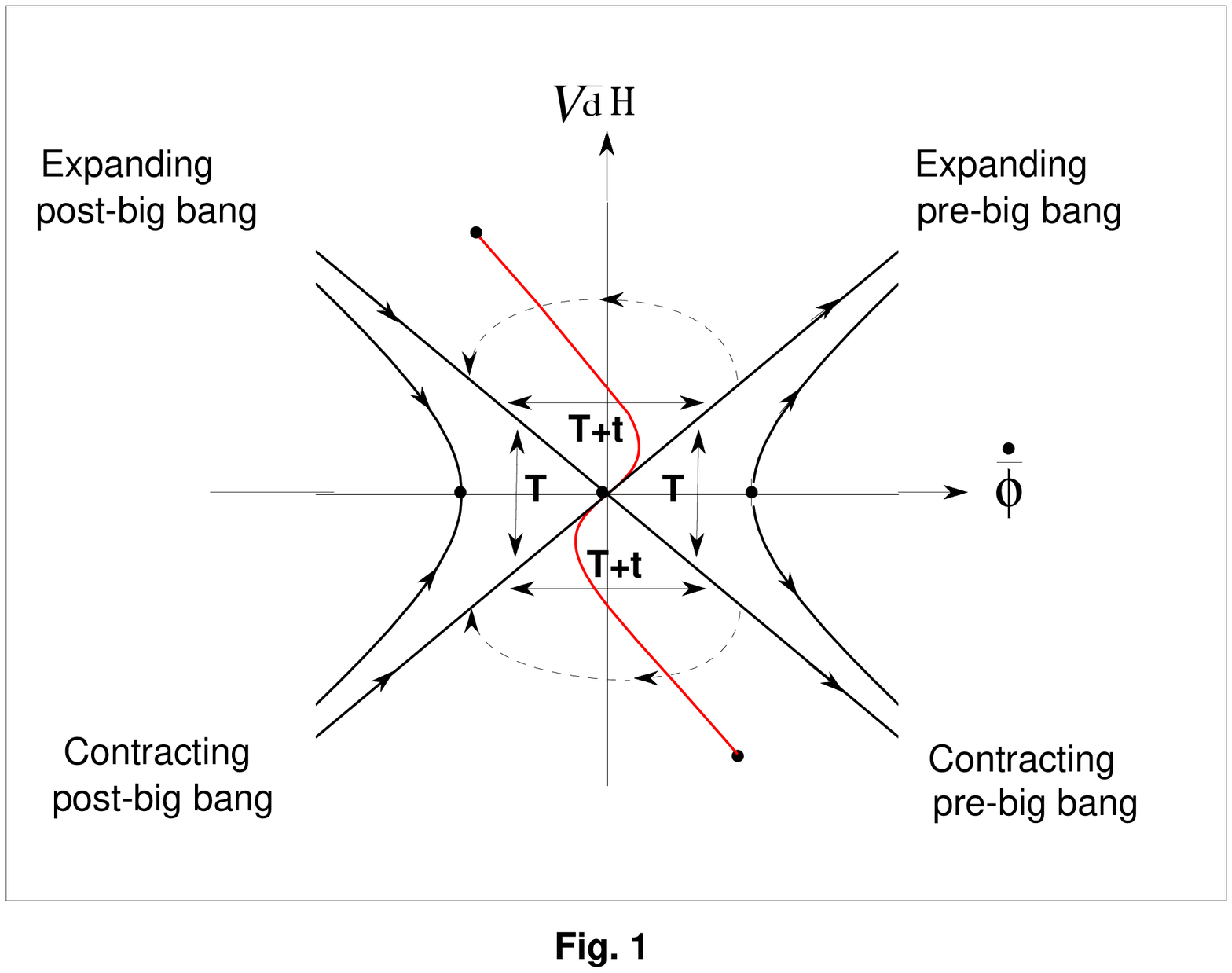}}
   \centerline{\parbox{11.5cm}{\caption{\label{fig:f1}
{\sl Cosmological evolution of the gravi-dilaton background according to
the string effective action, at tree-level in the string-loop 
expansion. }}}}
\end{figure}

The origin of the axes represents the perturbative vacuum, where
$H=0$ and $\phi=-\infty$. The bisecting lines represent the solutions of
the lowest-order effective action\cite{1,1a,1b}, 
with $V=0$ and without loops and
$\ap$ corrections. The lines in the upper-half plane
$H>0$ correspond to expanding configurations, in the lower half-plane
$H<0$ to contracting configurations. For $\fbp >0$ they describe an
accelerated evolution from the vacuum to the large curvature and
coupling ``stringy" regime (pre-big bang configurations); for $\fbp<0$
they describe a decelerated evolution from the stringy regime to the
vacuum (post-big bang configurations). The four branches of the
classical solutions are related by {\it T}-duality transformations
\beq
\fbp \Longleftrightarrow  \fbp ~~~~~~~~
H  \Longleftrightarrow -H , 
\label{14}
\eeq
and time-reversal transformations
\beq
\fbp \Longleftrightarrow - \fbp ~~~~~~~~
H  \Longleftrightarrow -H , 
\label{15}
\eeq
as indicated in Fig. 1. Both transformations are required\cite{3} for a
monotonic (expanding or contracting) transition from pre- to post-big
bang. In the case of a constant positive potential $V=\La >0$, the
solutions\cite{3} of the lowest-order action are represented by the
two branches of the hyperbola plotted in Fig. 1. The perturbative
vacuum is replaced in this case by a different configuration with flat
metric and linearly evolving dilaton, $\dot\phi = {\rm const}$, but the
symmetry pattern is preserved.

In the absence of higher-order corrections, the
four branches of the solutions are classically disconnected by a
curvature singularity that cannot be removed by any realistic choice
of the (local) dilaton potential\cite{4}. With an appropriate potential,
however, there are quantum transitions from pre- to post-big bang
that preserve the monotonic evolution of the scale factor, and which
are represented by the dashed curves of Fig. 1. The transition is
allowed, in particular, between two states\cite{5} with $\La=0$,
between two states\cite{6} with the same finite value of $\La$, and
also\cite{7} from the vacuum to a state with $\La >0$. These processes
are described by the scattering of the Wheeler--De Witt wave function
that represents our cosmological
configurations, in the two-dimensional minisuperspace spanned by the
coordinates $\fb=\phi-d \ln a$ and $\b=\sqrt d \ln a$. 

If we include the higher-derivative terms in the effective action,
however, the transition from the pre- to the post-big-bang sector of
Fig. 1 may become allowed even classically, and already to first
order in $\ap$. This effect is illustrated by the bold solid curve crossing
the origin in Fig. 1, which is obtained by numerically integrating the
equations following from the action (\ref{11}), without potential and
loop corrections, but with the four-derivative Gauss--Bonnet and
dilaton terms included\cite{8}. Starting from the perturbative vacuum,
it is found in that case that 
the Universe necessarily evolves towards a final
state with constant curvature, linearly running dilaton and $\fbp <0$, a
fixed point of the cosmological equations.

Assuming that, with the inclusion of the appropriate loop
corrections\cite{9}, the transition to the post-big bang phase is
successfully completed towards a state of radiation-dominated
evolution, we may wonder whether the phenomenological predictions
of such a string cosmology scenario are or not significantly different
from those of the standard scenario. The answer to this question is
known, and it is affirmative. There are, in particular, three main effects
worth mentioning: 1) the production of a relic graviton
background\cite{1,10,11,11a}, much stronger, at high frequency, that the
one expected in the context of standard inflationary models (so strong
that the associated energy density may be of the same order as the
entropy stored in the present CMB electromagnetic radiation\cite{12});
2) the production of a cosmic background of relic
dilatons\cite{1b,10,13}; 3) the amplification of the vacuum fluctuations
of the electromagnetic field, and the consequent production of
``seeds" for the galactic magnetic fields\cite{14}. 

The status of our
present knowledge about the graviton background has been reviewed
and discussed in a recent paper\cite{15}. In this paper I will thus
concentrate my discussion on the general properties of the relic dilaton
background. 

\renewcommand{\theequation}{2.\arabic{equation}}
\setcounter{equation}{0}
\section{Scalar perturbations and dilaton production}
\label{sec:2}
\noindent
In a  cosmological context there are various mechanisms of dilaton
production: particle collisions at high temperature, coherent
oscillations around the minimum of a scalar potential, amplification of
the vacuum fluctuations.  We shall concentrate here on the third
possibility because, even if the  temperature remains too low to allow
thermal production, and oscillations are avoided through a symmetry
that ensures the coincidence of the minima of the potential at early and
late times\cite{16}, quantum fluctuations cannot be eliminated; also, 
they may be expected to represent the dominant source\cite{17} when the
inflation scale is not smaller than about $10^{16}$ GeV, as in the
context of our scenario (another possible mechanism, dilaton radiation
from cosmic strings, has recently been discussed in Ref. [21]).

The production of dilatons through the amplification of the vacuum
fluctuations of the dilaton background\cite{1b}, $\phi \ra \phi
+\da \phi$, is very similar to the process of graviton production
through the amplification of tensor perturbations of the metric
background\cite{19}, $g_{\mu\nu} \ra g_{\mu\nu}
+\da h_{\mu\nu}$.  Unlike  the graviton case, however, the analysis of 
dilaton perturbations is complicated by their coupling to the
component of the metric perturbations, and to the perturbations of the
matter sources. The sources may even be absent in the pre-big bang
phase, but they are certainly not absent when the dilatons re-enter
the horizon, in the radiation and matter-dominated epoch. 

The general system of coupled perturbation equations, for the
background (\ref{13}) and for a perfect-fluid model of sources, can 
easily be written down in the Einstein frame, in the standard longitudinal
gauge\cite{20}, in terms of the variables $\psi$, $\chi$, $\da \r$, $\da
p$, $\da u_i$ defined by:
\bea
&&ds^2=a^2\left[d\eta^2\left(1+2\psi\right)-
\left(1-2\psi\right)dx_i^2\right] , ~~~~~~~~~~~~~ \da \phi =\chi, 
\nonumber \\
&& \da T_0^0=\da \r , ~~~~~~~~~
 \da T_i^j=-\da p \da_i^j , ~~~~~~~~~
 \da T_i^0=(p+\r)\da u_i/a . 
\label{21}
\eea
The vector-like equation\cite{1b} then obtained is:
\beq
Z_k''+2{a'\over a}{\cal A}Z_k' +\left(k^2{\cal B} + {\cal C} \right) Z_k 
=0, 
\label{22}
\eeq
where the components of the doublet $Z_k^{\dagger}=(\psi_k, \chi_k)$
are the Fourier modes of the metric and dilaton perturbations, and the
prime denotes differentiation with respect to the conformal time
$\eta$. The $2\times 2$, time-dependent mixing matrices
\beq
{\cal A}=\pmatrix{{1\over 2}(2d-3+d\ep), & -{1-\ep \over 4(d-1)}\b
\cr -(d-1)[{cd\over 2}(1-d\ep)+\b], & {1\over 2}(d-1)-{c\over
4}(1-d\ep)\b \cr}\eeq
\beq{\cal B}=\pmatrix{\ep, & 0\cr -c(1-d\ep)(d-1), & 1 \cr}\eeq
\beq{\cal C}=\pmatrix{2(d-2){a'' \over a} +(d-2)[d-4+d\ep + {1-\ep
\over 2(d-1)} \b^2](\H)^2, & 
{\ep+1\over 2(d-1)}{V'} a^2 \cr
cd(d-2)(\ep-\ga)[d(d-1)-{\b^2\over 2}](\H)^2 
+2(d-2){V'} a^2 , 
& [{V''} -{c\over 2}(1-d\ep)
{V'}] a^2  \cr}
\label{25}
\eeq
where $c=\sqrt{2/(d-1)}$ and $V'=\pa V/\pa \phi$, 
depend on the matter equation of state, $\ga(t)=p/\r$,
on the fluid model of perturbations, $\ep (t)= \da p/\da \r$, 
on the dilaton potential $V(\phi)$, and on the explicit background
solution through the parameter  $\b=\dot \phi/H$. Once eq. (\ref{22}) is
solved, for a given set of parameters $\{\b,\ga, \ep \}$,  there are two
additional independent equations\cite{1b} that determine the density and
velocity contrast $\da\r$ and $\da u$ in terms of $\psi$ and $\chi$:
\bea
&&\pa_i\left[2(d-1)\left(\H(d-2)\psi +\psi^\pr\right)-\chi \phi' \right]
=(\r+p)a\da u_i , \\
&&\nabla^2\psi -d\H\psi^\pr -\left[d(d-2)\left(\H\right)^2 -{d-2\over 2(d-1)}
\phi^{\pr 2} \right]\psi=\nonumber\\
&&={1\over 2(d-1)}
\left(\phi ' \chi^\pr +{\pa V\over \pa \phi} a^2\chi + a^2\da \r \right).
\label{27}
\eea

An exact computation of the scalar perturbation spectrum thus
requires an exact solution of the coupled equations (\ref{22}). Also, the
correct normalization of the spectrum needed to 
determine the
amplification of the quantum vacuum fluctuations,  requires the
knowledge of the normal modes of oscillation of the system {\sl
gravi-dilaton background} $\oplus$ {\sl fluid sources}, namely of the
variables that diagonalize the perturbed action, and satisfy canonical
commutation relations\cite{20,21}. Such variables are known for the
pure metric--fluid system\cite{22}, and for the pure metric--scalar field
system\cite{23}, but not for the complete system (\ref{21}), when the
dilaton is coupled to matter.

The amplification of the normalized vacuum fluctuation spectrum has
been determined\cite{1b}, up to now, for the simple transition from a
$d=3$, dilaton-dominated pre-big bang phase with negligible matter sources
($T_\mu^\nu=0=\da T_\mu^\nu$), to a radiation-dominated phase with
adiabatic fluid perturbations ($\ga=\ep=1/3$) and with the dilaton
frozen ($\b=0$) at the minimum of the non-perturbative potential ($\pa
V /\pa \phi=0$). In such a phase the perturbation equations
(\ref{22}) are decoupled, the canonical variables are known, and the
spectrum of scalar and dilaton perturbations (neglecting a possible
mass term $\pa^2 V/\pa \phi^2=m^2$) turns out to be the same as the
graviton spectrum, with a slope that is cubic\cite{1b} modulo
logarithmic corrections\cite{24}.

It should be stressed, however, that such a spectral distribution cannot
be extrapolated\cite{25} down to frequency scales re-entering the
horizon after equilibrium, since in the matter era ($p=0$) the
perturbation equations are no longer decoupled, even if the dilaton
background is frozen at the minimum of the potential. They remain
coupled not only in the longitudinal gauge, but also in the
uniform-curvature gauge\cite{26}, an off-diagonal gauge more
appropriate to scalar perturbations when growing modes are
present\cite{24}. In addition, a cubic slope cannot be extrapolated 
up to the maximum amplified frequency scale, as the
slope is expected to be different (in general flatter) in the highest
frequency band for which the first horizon crossing occurs not in the initial
dilaton-driven phase, but in the subsequent high-curvature string
phase\cite{8}. 

In this paper we shall discuss phenomenological bounds that apply to
the total integrated dilaton spectrum. As the spectrum is generally a
non-decreasing function of frequency (because the curvature scale is
non-decreasing in the pre-big bang epoch), we may restrict our
analysis to the high frequency sector of the spectrum, for which all
modes re-enter the horizon in the radiation era. In that range the
perturbation equations are decoupled, the canonical normalization in
known, and the spectral distribution of the energy density in the
relativistic regime $\om \gg m$ can be parametrized as\cite{1b}:
\bea
&&\Om_\chi(\om, t)={\om  \over \r_c}{d\r_\chi(\om, t)\over d\om}=
\Om_\ga (t)\left(H_1\over M_p\right)^2
\left(\om\over \om_1\right)^\da , \nonumber \\
&&\om<\om_1={H_1 a_1 \over a(t)} , ~~~~~~~ \da >0, 
~~~~~~~ H_1 \simeq M_s .
\label{28}
\eea

Here $\r_c=3M_p^2H^2(t)/8\pi$ is the critical density, $M_p$ is the
Planck mass, $\Om_\ga \simeq (H_1/H)^2(a_1/a)^4$ is the CMB
electromagnetic energy density in critical units; $H_1$ is the curvature
scale at the time $t_1$ of the inflation--radiation transition (which in
the present scenario is of the same order as the string mass scale
$M_s$); $\om_1$ is the maximal frequency of the spectrum undergoing
parametric amplification (I have used the fact that the peak value of the
dilaton spectrum has to be\cite{27,28} of the same order as the peak
of the graviton spectrum\cite{11a}). Finally, $\da$ is a growing spectral
index, whose exact value is at present unknown for the reasons
mentioned above. 

Fortunately, the bounds that we shall
consider here are only weakly dependent on $\da$, and become
completely $\da$-independent for $\da~\gaq~ 1$. In the following
section I will thus assume $\da~\gaq~ 1$ to simplify the discussion, but
the analysis can be easily extended\cite{1b,13} to any value of $\da$. 

\renewcommand{\theequation}{3.\arabic{equation}}
\setcounter{equation}{0}
\section{Phenomenological bounds}
\label{sec:3}
\noindent
The main bound on the background of relic dilatons follows from the
fact that the dilatons cannot be massless, because they are coupled
non-universally to macroscopic bulk matter\cite{29}, thus inducing an
effective violation of the equivalence principle in the macroscopic limit
of weak gravitational fields. This may be reconciled with the present
tests of the equivalence principle\cite{fis} 
if the range of the dilaton force is
smaller than about $1$ cm, i.e. for a dilaton mass
\beq
m~ \gaq~ 10^{-4}~ {\rm eV} . 
\label {31}
\eeq

Because of the mass, the produced dilatons tend to become
non-relativistic, as their proper momentum is red-shifted. When the
dominant mode $\om_1(t)$ becomes non-relativistic, the total
integrated energy evolves in time like $a^{-3}$: 
\beq
\Om_\chi=\int^{\om_1} {d\om\over \om}{m\over\om_1}\Om_\chi(\om,
t) \simeq {m M_s\over M_p^2}\left(M_s\over H\right)^2\left(a_1\over
a\right)^3 , ~~ \om_1(t) <m,
\label {32}
\eeq
and starts to grow in time with respect to $\Om_\ga \sim a^{-4}$. The
transition to the non-relativistic regime necessarily occurs before the
present epoch $t_0$, because the dominant mode has today a proper
wave number smaller than the dilaton mass\cite{1b}, 
$\om_1(t_0)\simeq (M_s/M_p)^{1/2} 10^{-4}$ eV $ <  m$. 

In the matter-dominated era, $\Om_\chi$ remains frozen at the
constant value $\Om_\chi(t_{eq})\simeq (m M_s/M_p^2)
(M_s/H_{eq})^2(a_1/a_{eq})^3$, where $H_{eq}\sim 10^6 H_0\sim
10^{-55}M_p$ is the curvature scale at the time of matter--radiation
equilibrium (I am discussing here an order-of-magnitude estimate, and I
will neglect  the dependence of the bounds on the precise value of the
present Hubble parameter $H_0$). 
By imposing $\Om_\chi(t_{eq})<1$, to avoid a Universe
overdominated by the coherent oscillations of the produced
dilatons\cite{31}, we obtain the bound
\beq
m~\laq~ \left(H_{eq}M_p^4/ M_s^3\right)^{1/2} , 
\label{33}
\eeq
which represents, in our context, the most restrictive upper bound
if dilatons are not yet decayed. For $100$ keV $\laq ~m ~ \laq ~100$ MeV
a more restrictive constraint on $\Om_\chi$ is provided by the
observations of the diffuse $\ga$-ray background\cite{18}, but this
range of masses is excluded, in our case, by the allowed range of $M_s$ 
(see next section). 

Values of the dilaton mass higher than allowed by the  critical
density bound (\ref{33}) can be reconciled with present observations,
only if the energy stored in the coherent oscillations was dissipated
into radiation before the present epoch, at the decay scale $H_d>H_0$,  
fixed by the decay rate $\Ga_d$ of dilatons into photons, $H_d\simeq
\Ga_d \simeq m^3/M_p^2$. 
The reheating associated to this decay leads to an entropy increase 
$\Da S \simeq (T_r/T_d)^3$, where $T_r \simeq (M_p H_d)^{1/2}$ is the
final reheating temperature, and $T_d$ is the  radiation
temperature immediately before dilaton decay. This increase is
significant ($\Da S >1$) provided dilatons decay when they are
dominant, namely for $t>t_i$, where $t_i$ is the time scale marking the
beginning of dilaton dominance, $\Om_\chi(t_{i})=\Om_\ga(t_{i})$. From
eq. (\ref{32}) we have $H_i\simeq m^2M_s^3/M_p^4$ so that, for $H_d<H_i$, 
the radiation temperature before decay is $T_d=T_i (a_i/a_d)\simeq
(H_i M_p)^{1/2}(H_d/H_i)^{2/3}\simeq (m^{10}/M_s^3 M_p)^{1/6}$,
corresponding to an entropy increase
\beq
\Da S \simeq \left (M_s^3/ m M_p^2\right)^{1/2}.
\label{34}
\eeq
This entropy injection can in principle disturb nucleosynthesis or
baryogenesis\cite{31}, and we must consider two possibilities. 
\begin{itemize}
\item If the reheating temperature $T_r$ is too low to allow
nucleosynthesis, i.e. $m~\laq ~10$ TeV, we must assume that
nucleosynthesis occurred before, and we must impose $\Da S ~\laq
~10$ to avoid destroying the light nuclei already formed. A more
precise bound can be determined through a detailed analysis of
photodissociation\cite{32} and hadroproduction\cite{33} processes, but
such an increase of precision is irrelevant in our context since, as we
shall see, it refers to values of $m$ outside the allowed range. 
\item If the reheating temperature is  large enough to allow
nucleosynthesis, i.e. $m~\gaq ~ 10$ TeV, the only possible constraint
comes from primordial baryogenesis. The bound is model-dependent,
but the constraint $\Da S ~\laq ~10^5$ seems to be sufficient\cite{31} 
not to wash out any pre-existing baryon--antibaryon asymmetry
(this bound could be evaded in the case of low-energy baryogenesis,
occurring at a scale $H<H_d$). 
\end{itemize}

The previous bounds refer to the case $m<M_s$. If $m>M_s$ then the
produced dilatons are non-relativistic already from the beginning, their
total integrated energy density is\cite{1b}
\beq
\Om_\chi(t) \simeq \left(m\over M_p\right)^2
\left(M_s\over H\right)^2\left(a_1\over a\right)^3 , 
\label {35}
\eeq
and the only bound to be imposed is $m<M_p$, to avoid overcritical
density. There are no additional bounds, as the dilatons decay before
becoming dominant. 

\renewcommand{\theequation}{4.\arabic{equation}}
\setcounter{equation}{0}
\section{Allowed mass windows}
\label{sec:4}
\noindent
By intersecting the region allowed by the previous phenomenological
bounds, with the allowed values of the string mass scale\cite{34}, 
$0.01~\laq~ M_s/M_p~\laq ~0.1$, we obtain for the dilaton mass the
two windows illustrated in Fig. 2:
\beq
10^{-4}~ {\rm eV} ~\laq ~m ~ \laq ~ 10~ {\rm keV}, ~~~~~~~~~~~~~~
10~ {\rm TeV} ~\laq ~m .
\label{41}
\eeq
\begin{figure}[htb]
   \epsfxsize=11cm
   \centerline{\epsfbox{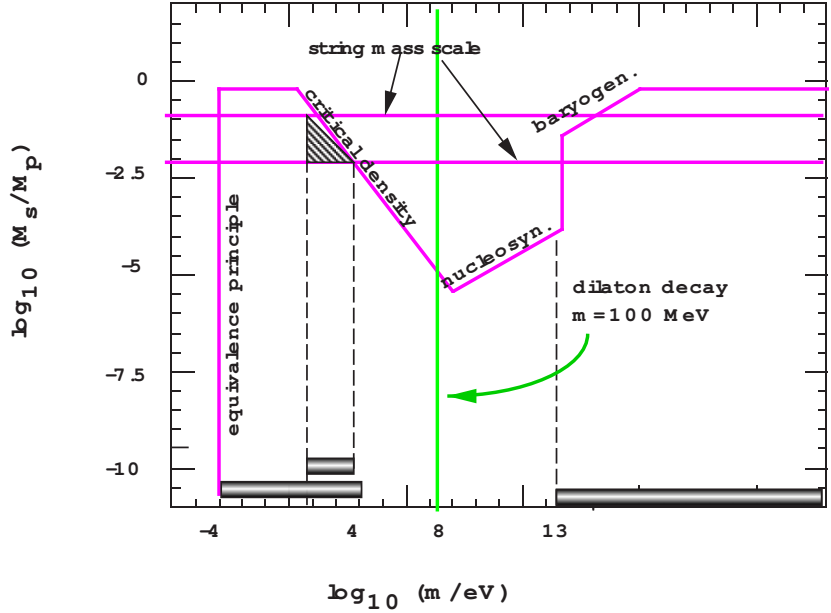}}
   \centerline{\parbox{11.5cm}{\caption{\label{fig:f2}
{\sl Allowed dilaton mass windows for a spectral slope $\da \geq 1$.
The shaded triangle defines the region of parameter space compatible
with a dominant contribution of relic dilatons to the present critical
density. }}}} 
\end{figure}
They lie on the opposite sides of the $100$ MeV decay line,
corresponding to a dilaton 
decay scale of the same order as our present Hubble
scale, $H_d \simeq H_0$. If the mass is in the right window then all
produced dilatons have already decayed, and no background is
available today to direct observation. If, on the contrary, the mass is
inside the left window, then the background is still around us and could
be observed, in principle, by those experimental devices that are
sensitive to scalar oscillations of gravitational strength, 
such as spherical gravity wave detectors\cite{35}. Unfortunately,
since dilatons today are non-relativistic, they should oscillate
coherently at a frequency determined by their rest mass, and thus
higher than about $100$ GHz, according to eq. (\ref{31}). This is clearly
outside the typical frequency range of such detectors. 
Resonant microwave cavities, used in the search of cosmic
axions\cite{35a}, are  also  disfavoured because of the very small
coupling of dilatons to photons, which is at least a factor $10^{-8}$
smaller than the corresponding coupling of axions.   

In spite of the fact that a dilaton mass in the TeV range seems to be at
present supported by supersymmetry-breaking motivations\cite{36}, it
is important to recall that a dilaton mass in the left window is all but
theoretically excluded, as shown for instance by models of
supersymmetry breaking with light dilatons\cite{37}. It is also worth
stressing that, in the restricted range
\beq
100~ {\rm eV} ~\laq~ m ~\laq ~ 10~{\rm keV},
\label{42}
\eeq
the dilatons could saturate the critical density bound, as illustrated
in Fig. 2, thus becoming an attractive dark matter 
candidate\cite{1b,13} 
with $0.01 ~\laq ~ \Om_\chi ~\laq~1$ according to eq. (\ref{32}). 

The  mass windows of Fig. 2 refer to a dilaton spectrum that grows
linearly or faster with frequency, $\da \geq 1$. If $ \da <1$ 
the bounds 
become $\da$-dependent and slightly more constraining\cite{1b,13},
and the allowed windows are further reduced. The critical density
bound (\ref{33}), in particular, becomes
\beq
m~\laq~\left(H_{eq}M_p^4M_s^{\da-4}\right)^{1/(\da+1)}, ~~~~~~~
\da \leq 1,
\label{43}
\eeq
and the left window disappears completely for $\da~\laq~ 0.5$.

In this sense, the mass windows of eq. (\ref{41}) represent the
maximally extended allowed range for the dilaton mass, at least in the
context of a ``minimal" pre-big bang scenario\cite{15} in which the
CMB radiation that we observe today is entirely produced at the end of
the string phase at a scale $H_1 \simeq M_s$. It is not impossible,
however, to imagine  more complicated (but perhaps also more
unnatural) models in which the phenomenological bounds determining
the allowed mass window are relaxed, because of an additional
reheating phase occurring before nucleosynthesis, and before the
beginning of dilaton dominance. Such a reheating could be the
consequence of a phase of ``intermediate scale" inflation\cite{38} or
of ``thermal" inflation\cite{39}, and is only constrained by the
requirement of a negligible dilution of any pre-existing baryon number
(but baryogenesis could be even produced by the ``flaton"
field\cite{39} itself, whose decay is responsible for the additional 
reheating). 

Any mechanism producing a significant amount of thermal radiation
(associated or not to a phase of inflation) indeed dilutes  the original
dilaton density (\ref{28}), with respect to $\Om_\ga$, by the
factor\cite{11a}
\beq
\Om_\chi \ra \Om_\chi(1-\da s)^{4/3}\left(n_f/ n_b\right)^{4/3}, 
~~~~~ \da s ={(s_f-s_b)/s_f} . 
\label{44}
\eeq
Here $s_b, s_f, n_b, n_f$ are, respectively, the thermal entropy density
of the CMB radiation and the number of particles species in thermal
equilibrium, at the beginning ($t_b$) and at the end ($t_f$) of the
reheating process. An efficient reheating, $s_f \gg s_b$, $\da s\ra 1$,
reduces in a significant way the dilaton fraction of critical
density $\Om_\chi$: as a consequence, the scale of dilaton dominance
$H_i$ is lowered, the decay temperature $T_d$ is raised, and the
bounds on $m$ following from the entropy constraint $\Da S <10$ and
the critical bound $\Om_\chi<1$ are relaxed. We can easily estimate, by
assuming in particular $n_f\sim n_b$, that the mass gap between the
left and right windows of eq. ({\ref{41}) is completely filled for
\beq
1-\da s ~\laq ~ 10^{-4} ,
\label{45}
\eeq
namely for an intermediate reheating phase producing more than
$99.99$\% of the entropy at present stored in the thermal black-body
background. 

An appropriate reheating process can thus easily render 
a dilaton mass of the TeV order compatible 
with the string inflation scale $M_s$. 
The final allowed region should not be further reduced by other
bounds applied to additional processes of production since, for an
inflation scale of the order of $M_s$, the amplification of the vacuum
fluctuations is expected to represent the dominant 
mechanism\cite{17} of dilaton production. 

\renewcommand{\theequation}{5.\arabic{equation}}
\setcounter{equation}{0}
\section{Conclusion}
\label{sec:5}
\noindent
The evolution from the string perturbative vacuum to the radiation era,
described by pre-big bang models of the early Universe, is accompanied
by the parametric amplification of the quantum fluctuations of the
dilaton background, and leads to the production of a sea of relic cosmic
dilatons. Their spectral distribution is in general non-decreasing with
frequency, and is normalized to a peak value determined by the string
mass scale, such as that of the relic graviton spectrum.

Our present ignorance of the kinematic details of the model at high
curvature scale, and the complicated mixing with matter at the time of
re-enter, have prevented so far a definite prediction for the spectral
index, in both  the high and the low frequency sectors. Using however the
general properties of the spectrum, we can obtain a reliable estimate
of the allowed range for the dilaton mass. The extension of such range
depends only weakly on the unknown spectral slope, and becomes
completely slope-independent for a spectrum that grows linearly or
faster.

The analysis performed has produced, in particular, the following two 
results. 1) The allowed mass windows may include a range of values in
which the cosmic dilatons are not yet decayed, and provide a dominant
contribution to the present critical energy density. 2) With the
introduction of an intermediate reheating stage, subsequent to the
string-radiation transition, a non-minimal model can easily be made
compatible with a dilaton mass in the TeV range, the present preferred
value of the conventional supersymmetry-breaking scenario.

\vspace{1cm}
{\it Acknowledgements:\/} I am grateful to  Gabriele Veneziano for
a fruitful and enjoyable collaboration on the pre-big bang scenario and,
in particular, on the associated background of relic dilatons. It is 
a pleasure to thank also Emilio Picasso for interesting discussions and
useful information on resonant microwave cavities.

\end{document}